\documentclass[aps,prb,twocolumn]{revtex4-2}
\usepackage{graphicx}
\usepackage{bm}
\usepackage{hyperref}

\begin{document}

\title{Domain–Direct Band Gaps: Classification and Material Realization}

\author{Yalan Wei$^1$,$^\#$}
\author{Hairui Ding$^2$,$^\#$}
\author{Shifang Li$^3$}
\author{Yuke Song$^1$}
\author{Chi Ren$^2$}
\author{Xiao Dong$^2$}
\email{xiao.dong@nankai.edu.cn}
\author{Chaoyu He$^1$$^3$}
\email{hechaoyu@xtu.edu.cn}
\affiliation{$^1$School of Physics and Optoelectronics, Xiangtan University, Xiangtan 411105, China}
\affiliation{$^2$School of Physics and MOE Key Laboratory of Weak-Light Nonlinear Photonics, Nankai University, Tianjin, 300071, China}
\affiliation{$^3$Center for Quantum Science and Technology, Shanghai University, Shanghai 200444, China}

\begin{abstract}
The conventional classification of direct band-gap semiconductors relies on point-like extrema in momentum space. Here, we introduce the concept of domain-direct band gaps, where the conduction-band minimum (CBM) and valence-band maximum (VBM) form extended manifolds in the Brillouin zone. We demonstrate this concept through the material realization of an extreme two-dimensional–two-dimensional (2D–2D) domain-direct band gap in twisted diamond. First-principles calculations show that both the CBM and VBM exhibit nearly flat 2D manifolds in the \( k_x - k_y \) plane with minimal energy variation (a few meV), yielding a direct band gap of 3.264 eV. In contrast, strong dispersion along the out-of-plane \( k_z \) direction induces anisotropic carrier dynamics, with strongly suppressed in-plane Fermi velocities (down to \( \sim 10^1 - 10^3 \, \text{m/s} \) in certain directions) and much larger out-of-plane velocities (\( \sim 10^6 \, \text{m/s} \)). The nearly flat CBM and VBM manifolds enhance the joint density of states, leading to a pronounced optical absorption peak at the band gap onset. This new type of domain-direct gap, coupled with strong directional anisotropy, opens up opportunities for anisotropic optoelectronic applications. Our results establish domain-direct band gaps as a new class of semiconductors, demonstrating their feasibility in real materials.
\end{abstract}
\maketitle

In crystalline solids, the distinction between direct and indirect band gaps plays a central role in determining optical and transport properties, particularly in semiconductors and insulators. Within the conventional framework of band theory, a band gap is termed direct when the valence-band maximum (VBM) and the conduction-band minimum (CBM) occur at the same crystal momentum, and indirect when they occur at different momenta. This classification, which underpins the design of optoelectronic materials, implicitly assumes that both the VBM and CBM are isolated extrema at discrete points (0D) in momentum space [1–3]. While this assumption is well justified for many conventional semiconductors, it is not guaranteed in more complex band structures featuring degeneracies, reduced dispersion, or symmetry-protected manifolds.

Recent advances in electronic structure theory and materials discovery have revealed a wide range of systems in which band extrema are not point-like objects, but instead form extended structures in momentum space. Representative examples include flat or nearly flat bands [4-11] and semimetals with zero-energy nodal line [12-14], where the energy dispersion is quenched or highly anisotropic along certain momentum directions [15-17]. In such cases, the valence-band top or conduction-band bottom may correspond to a one-dimensional (1D) line, two-dimensional (2D) surface, or three-dimensional (3D) manifold in reciprocal space, rather than isolated k-points. Notably, flat bands — characterized by an extremely small bandwidth and a diverging density of states — have attracted intense interest due to their propensity for correlation-driven phenomena, including magnetism and superconductivity [18-20]. These developments expose a conceptual gap in the traditional direct/indirect classification, which does not distinguish between qualitatively different types of momentum-space alignment between extended band extrema.

Motivated by these observations, it is natural to generalize the notion of a direct band gap by explicitly considering the geometric dimensionality of the band extrema in momentum space. Beyond the conventional point–point direct gap, one may identify 16 distinct classes of direct band gap semiconductors, depending on whether the VBM and CBM are localized at points, lines, surfaces or volumes in reciprocal space. Building on this framework, we explore twisted diamond structures and identify a 2D–2D domain-direct band gap, in which both the CBM and VBM form nearly flat 2D manifolds in the kx-ky plane. These flat manifolds, combined with pronounced dispersion along kz, lead to enhanced joint density of states and strongly anisotropic carrier dynamics, providing a realistic example of how extended band extrema can generate novel electronic and optical behavior in three-dimensional materials.

\begin{figure*}[t]
\centering
\includegraphics[width=0.9\linewidth]{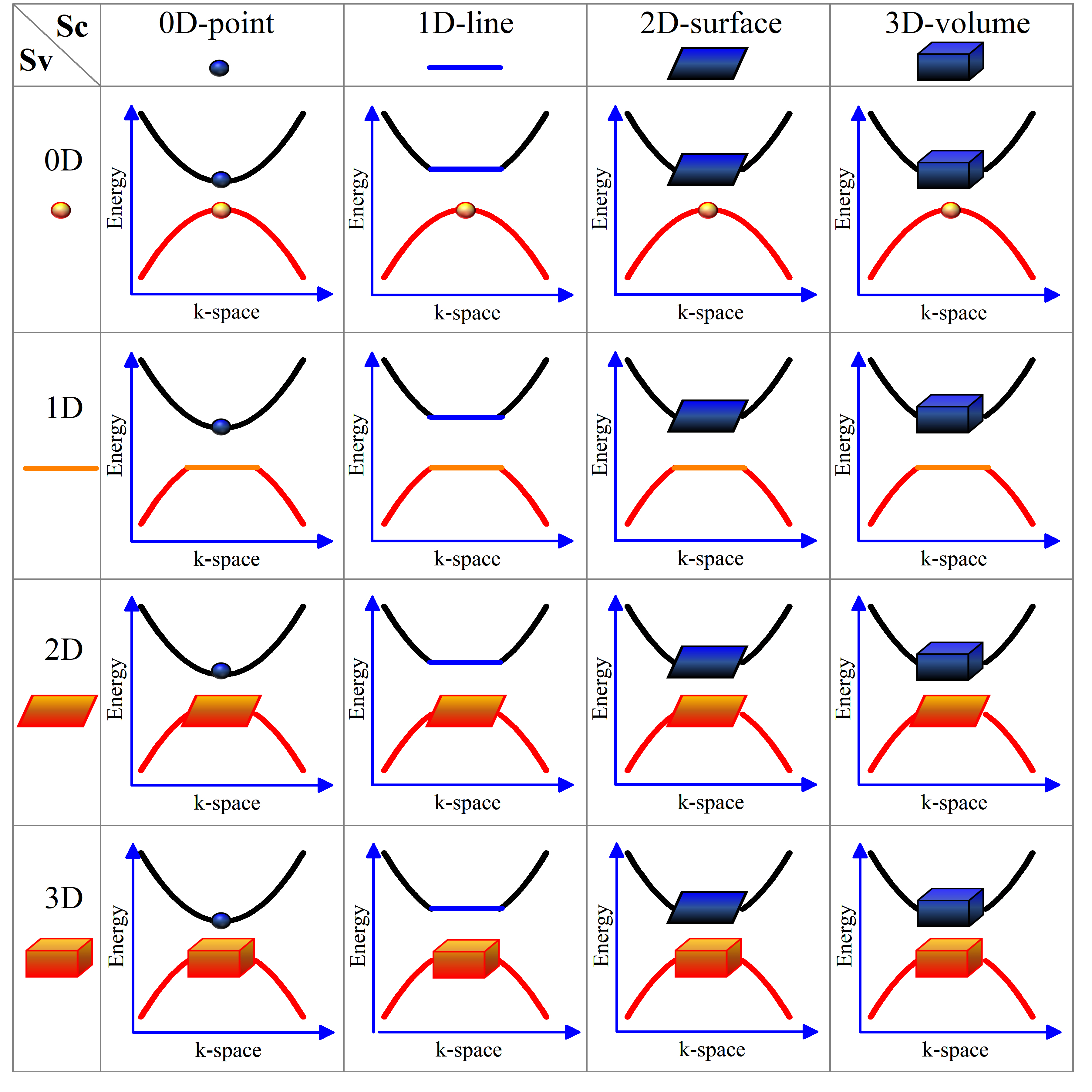}
\caption{Generalized classification of direct band gaps based on the geometry of their band-extremum sets in k-space.}
\end{figure*}

The generalized classification of direct band gap semiconductors is based on the geometry of their band-extremum sets in k-space. We use Sv and Sc to denote the sets of k-points corresponding to the VBM and CBM, respectively. Each set may possess distinct dimensionality: 0D (point), 1D (line), 2D (surface) and 3D (volume). If Sv and Sc have an intersection, the material is classified as a direct band-gap semiconductor. When the intersection consists of only discrete points, it corresponds to a conventional point-to-point direct gap; when the intersection forms a continuous region, it can be defined as a domain-direct gap. As shown in Fig. 1, all 16 possible types of direct band gaps can be classified, including 0D-0D, 0D-1D/1D-0D 0D-2D/2D-0D, 0D-3D/3D-0D, 1D-1D, 1D-2D/2D-1D, 1D-3D/3D-1D, 2D-2D, 2D-3D/3D-2D and 3D-3D. Here, 0D, 1D, 2D, and 3D denote isolated points, lines, surfaces, and extended volumes of extrema in k-space, respectively. Each combination corresponds to a distinct geometric relation between the valence-band maximum (VBM) and conduction-band minimum (CBM), forming a complete framework for conventional point-like direct gaps as well as extended domain-direct gaps.

\begin{figure*}[t]
\centering
\includegraphics[width=0.9\linewidth]{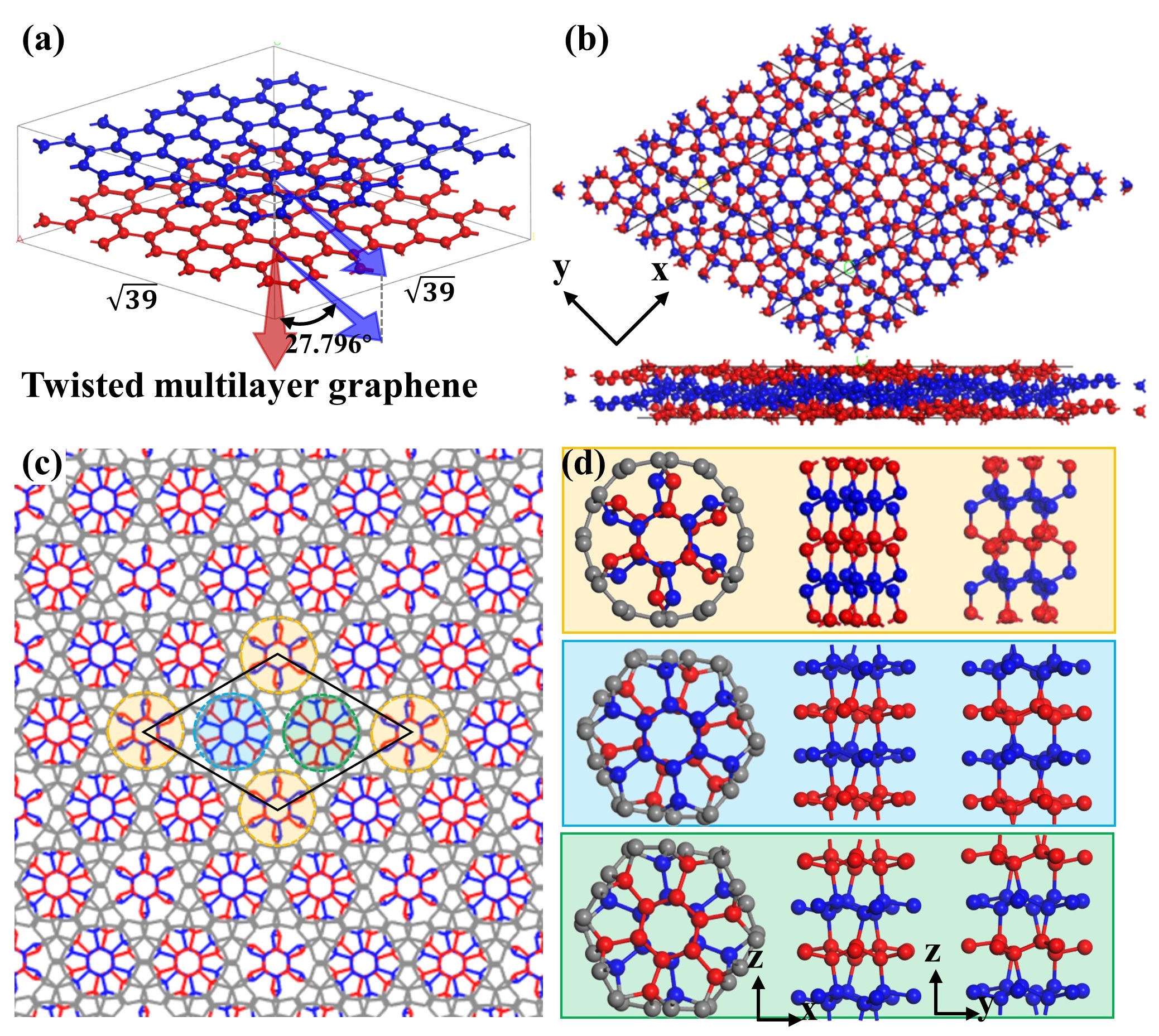}
\caption{(a) Top view of the twisted bilayer graphite with a $\sqrt{39}$×$\sqrt{39}$ supercell at a twist angle of 27.8°. (b) Top and side views of the twisted diamond obtained after introducing interlayer hybridization, where different layers are indicated by different colors. (c) Top view highlighting the formation of three nanotube-like backbones arranged in a triangular pattern as a result of interlayer hybridization. (d) Enlarged views for detailed atomic configurations on the nanotube-like backbones.}
\end{figure*}

A domain-centric perspective highlights clear advantages of extended or overlapping band extrema, including enhanced optical phase space, increased joint density of states, and enhanced anisotropy of electronic and optical responses. In particular, flat 2D manifolds at the band edge provide platforms for novel excitonic, orbitronic, and strongly correlated phenomena, enabling potential applications such as multi-state information storage and isotropic or anisotropic optoelectronics. However, the formation of higher-dimensional extrema is increasingly constrained by symmetry requirements. The 3D flat bands—where the VBM and CBM span the entire Brillouin zone (BZ) volume—are extremely rare due to stringent symmetry constraints. This scarcity naturally motivates a focus on 2D extrema, where VBM and CBM form extended surfaces in k-space, which have been successfully realized in materials such as twisted bilayer graphene [21-23], hexagonal boron nitride (h-BN) [24], and transition metal dichalcogenides (e.g., MoS2) [25-27]. Through layer twisting, strain engineering, or interlayer potential modulation, these systems achieve nearly flat conduction or valence bands, with VBM and CBM forming contiguous 2D manifolds or quasi-flat regions, demonstrating practical routes to engineer domain-direct behavior. Inspired by this “twist engineering” concept, we further explored a series of twisted diamond structures and identified 2D–2D type domain-direct semiconductors among them.

To demonstrate the feasibility of domain-direct band gaps in realistic materials, we consider a 3D carbon system—twisted diamond—as a representative example. The structure is generated by introducing interlayer hybridization, using the graph-based crystal structure prediction method RG2 [28], into a twisted graphite phase constructed from a $\sqrt{39}$×$\sqrt{39}$ supercell with a twist angle of 27.80°, as shown in Fig. 2(a). Despite the substantial bond reconstruction associated with the sp²–sp³ transition at the complex interface, the optimized structure retains a relatively high overall symmetry and belongs to space group No. 165 (P-3m1). Owing to the large moiré periodicity induced by twisting, the system contains 13 inequivalent atomic sites (156 total atoms) and a variety of locally distinct bonding environments (5-, 6- and 7-member rings). We label this structure as 165-13-156-r567-0 according to these features. In Fig. 2(b), we present only the top and side views of the optimized crystal structure. Atoms originating from the two graphite layers are highlighted in red and blue, respectively, clearly revealing the moiré pattern induced by the twist, as well as the emergence of three nanotube-like backbones extending along the out-of-plane z-direction in Fig. 2(c-d). These quasi-one-dimensional units are periodically arranged at the special positions (0, 0), (1/3, 2/3), and (2/3, 1/3) of the hexagonal lattice, forming a triangular lattice. The full structural information is provided in the Supplementary Materials in table S1 for allowing interested readers to inspect the structure in detail.

\begin{figure*}[t]
\centering
\includegraphics[width=0.9\linewidth]{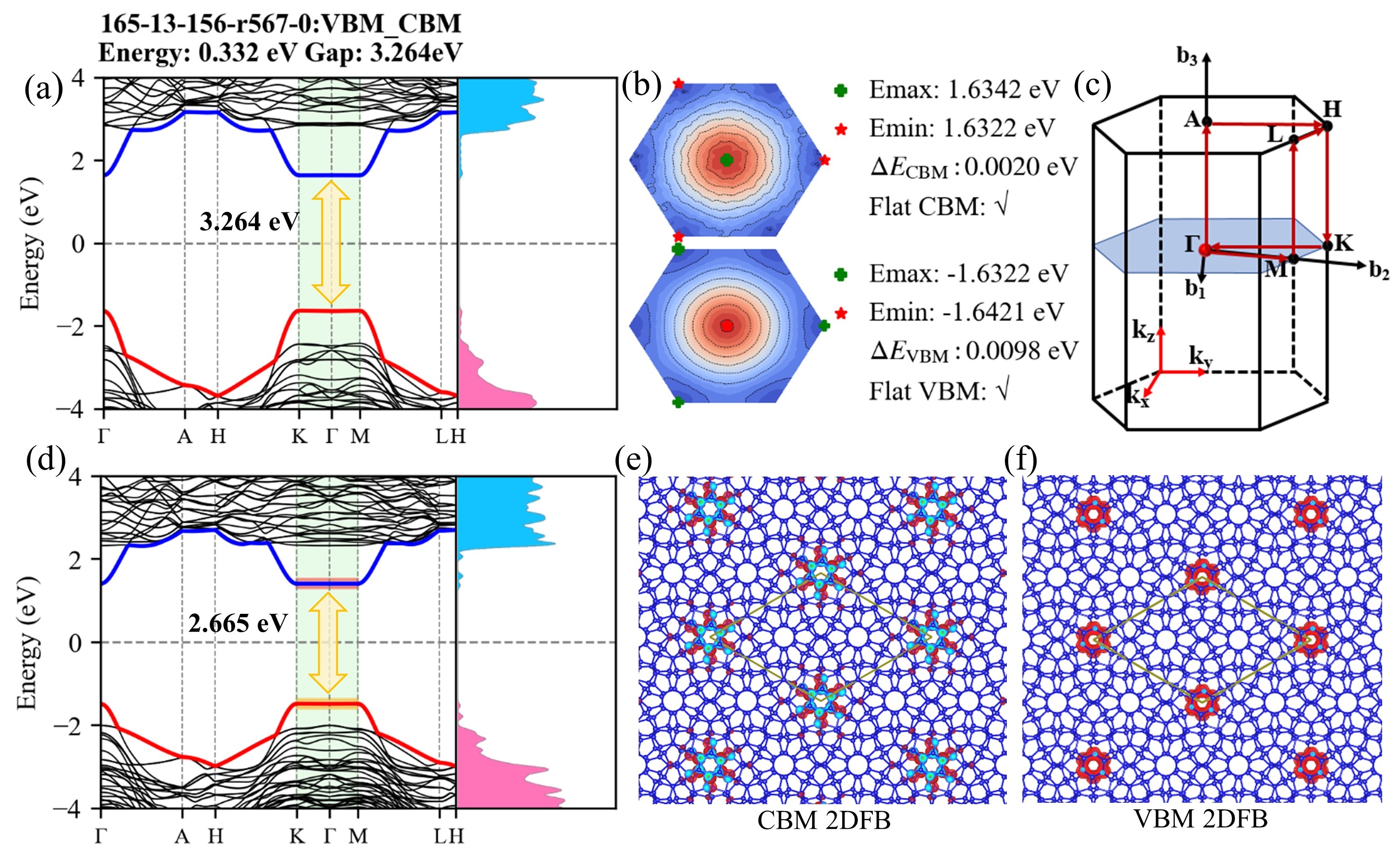}
\caption{ (a) TB-based band structures and total density of states (DOS). (b) 3D band structures of the highest valence band and the lowest conduction band in the kz = 0 plane of the 2D BZ (green-filled region). (c) The conventional high-symmetry paths for a 3D hexagonal lattice. (d) PBE-based band structures and total density of states (DOS). (e) Real-space partial charge density distribution corresponding to the conduction band minimum flat band. (f) Real-space partial charge density distribution corresponding to the valence band maximum flat band.}
\end{figure*}

Energetically, the twisted diamond structure has an average energy of approximately -8.756 eV/atom, which is higher than that of bulk diamond by 0.332 eV/atom. Although it is not the thermodynamic ground state of carbon, this energy difference does not imply intrinsic instability. Compared with the corresponding twisted bilayer graphene structure, the twisted diamond phase is higher in energy by about 0.455 eV/atom, reflecting the energetic cost associated with the transition from sp² to sp³ bonding. More importantly, the twisted diamond structure is found to be dynamically stable. As shown in Fig. S1, the calculated phonon spectrum shows no imaginary frequencies throughout the BZ, indicating dynamical stability to against small vibration. In addition, first-principles molecular dynamics simulations (MD) confirm its finite-temperature robustness: throughout the simulation, the interlayer sp³-hybridized framework remains intact, with no tendency to revert to twisted graphite or other sp²-bonded configurations. This demonstrates that once formed, the twisted diamond structure is kinetically locked, providing a solid and reliable platform for investigating its unconventional electronic and optical properties.

\begin{figure*}[t]
\centering
\includegraphics[width=0.9\linewidth]{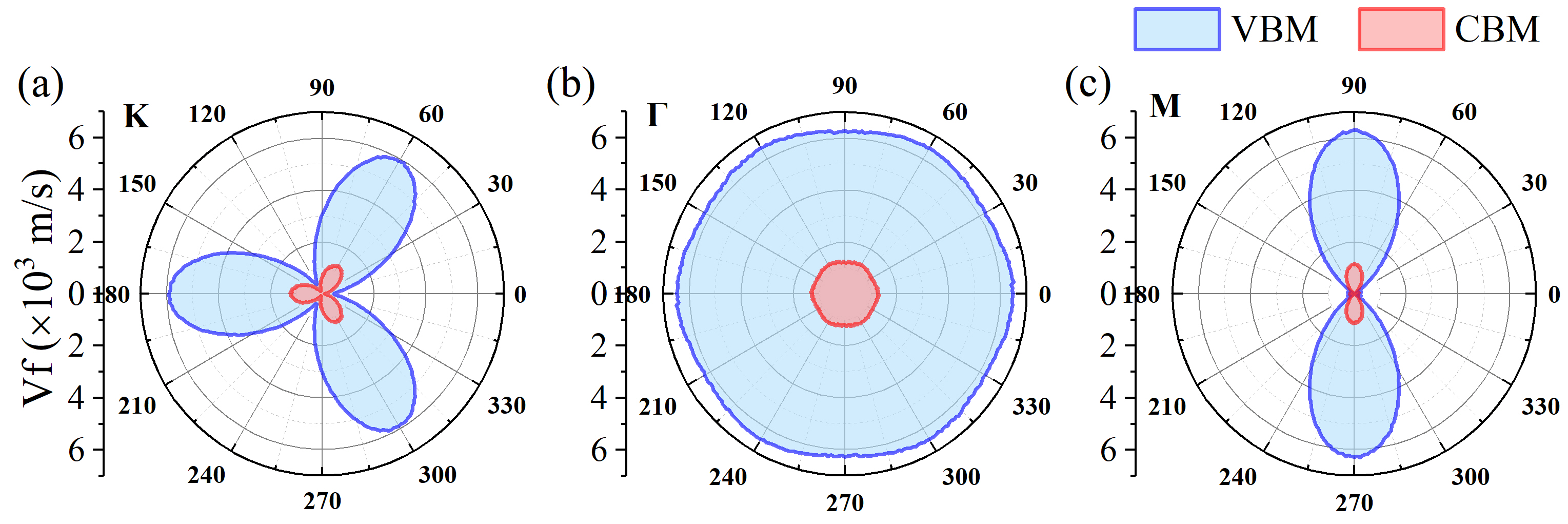}
\caption{In-plane projections of the direction-dependent Fermi velocities for the VBM (blue) and CBM (red) at the (a) K, (b) $\Gamma$, and (c) M points in kz=0 plane.}
\end{figure*}

Owing to the large unit cell containing 156 atoms, performing fully self-consistent HSE calculations is computationally prohibitive. Instead, we employ a gt-TB model [29] with carbon parameters benchmarked against HSE accuracy, whose reliability and transferability across diverse carbon allotropes have been systematically validated in previous work [30]. Using this gt-TB model, we first calculate the band structure along the conventional high-symmetry paths of the folded hexagonal BZ, as shown in Fig. 3 (a) and (c). A striking feature is that both the VBM and CBM occur within the 2D BZ and exhibit nearly dispersionless behavior, appearing as almost flat lines along the high-symmetry directions. The resulting band gap is approximately 3.264 eV, characteristic of a wide-gap semiconductor. To further quantify the flatness of the band edges, we perform high-density sampling of the entire 2D BZ at kz=0, as shown in Fig. 3(b). The energy variation of the VBM is found to be about 9.8 meV, while that of the CBM is even smaller, approximately 2.0 meV, confirming the formation of nearly ideal 2D band-edge manifolds. As an independent cross-check, we also calculate the band structure of twisted diamond along the same high-symmetry paths using VASP within the PBE functional. As shown in Fig. 3 (d), the overall band dispersions are in excellent qualitative agreement with those obtained from the gt-TB model, while the separation between the VBM and CBM is reduced, yielding a smaller band gap of 2.665 eV. This behavior is fully consistent with the well-known tendency of PBE to underestimate band gaps relative to HSE, thereby providing further support for the reliability of the gt-TB description. 

Finally, based on the PBE wave functions, we analyze the real-space charge density distributions associated with the CBM and VBM, as shown in Fig. 3 (e) and (f). Both states are found to be predominantly localized on the same sublattice of the triangular network formed by the nanotube-like motifs, exhibiting pronounced localization in the x–y plane while remaining extended along the z direction. This spatial anisotropy provides a real-space counterpart to the strongly anisotropic electronic dispersion in momentum space, offering physical insight into the origin of the nearly flat band dispersion at the band edges. To further examine the dimensionality of the flat bands, we sampled the band dispersion on a series of 2D kz-resolved planes with $k_z = 0.1, 0.2, 0.3, 0.4, \text{ and } 0.5 \, (\text{in units of } 2\pi/c)$, as presented in Fig. S2. We find that the CBM remains remarkably flat within each kz plane, with an in-plane bandwidth consistently below 20 meV over the entire kz range considered. In contrast, the VBM exhibits a rapidly increasing in-plane dispersion as kz increases: while its bandwidth is limited to approximately 10 meV at kz = 0, it grows continuously and reaches values as large as ~250 meV at kz = 0.5. These results reveal a pronounced asymmetry between the conduction and valence band edges.

\begin{figure*}[t]
\centering
\includegraphics[width=0.9\linewidth]{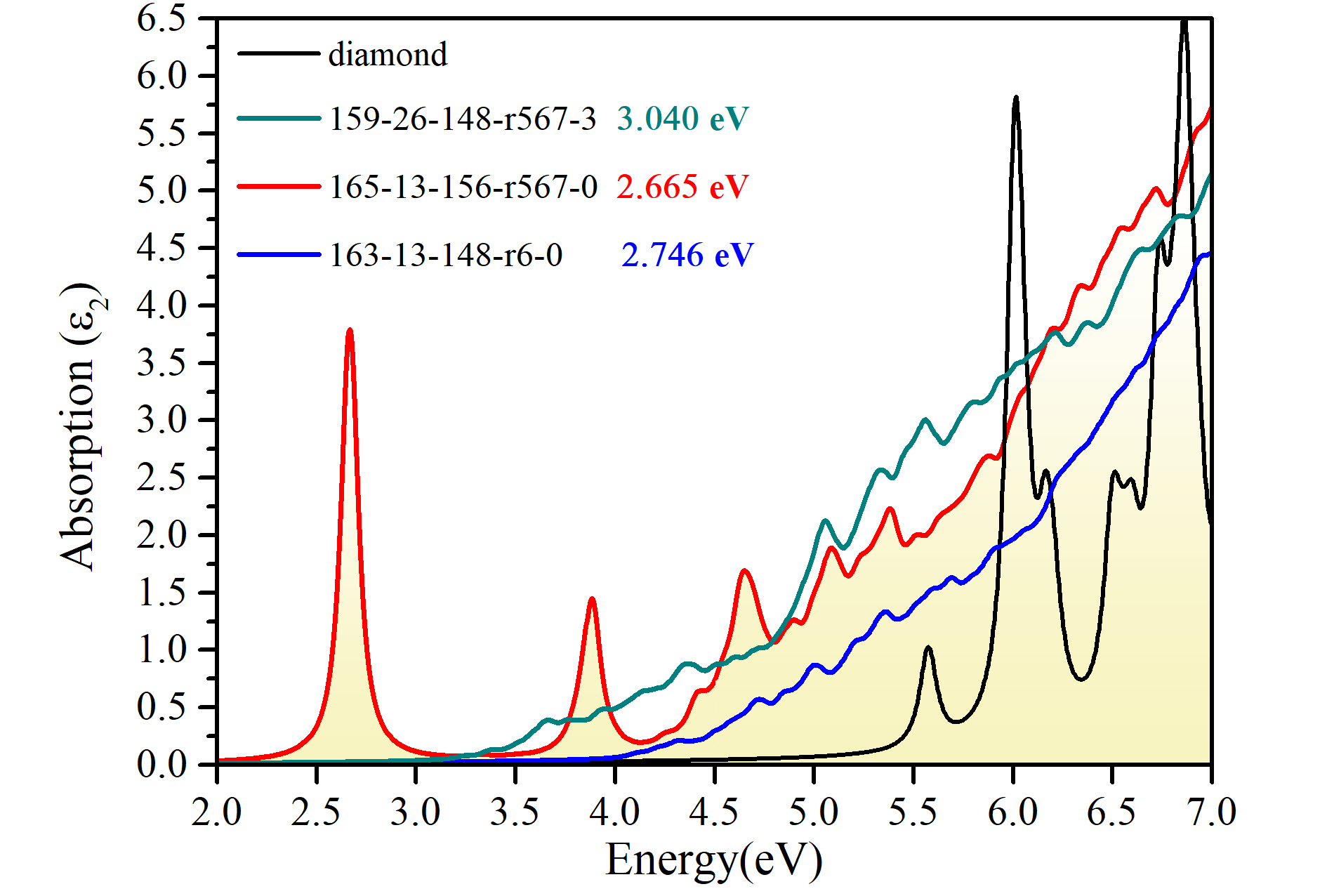}
\caption{PBE-calculated optical absorption spectra of diamond and the large-supercell structures 165-13-156-r567-0, 163-13-148-r6-0, and 159-26-148-r567-3. The corresponding PBE band gaps are indicated for each structure and their bands structures are shown in Fig.S16.}
\end{figure*}

To further elucidate the anisotropic nature of the domain-direct band edges, we evaluate the direction-dependent Fermi velocities of both electrons and holes at the K, $\Gamma$, and M points within the kx–ky plane under kz = 0, as shown in Fig. S3. The corresponding in-plane projections are presented in Fig. 4, with the extrema summarized in Table S2. Significant differences in both magnitude and anisotropy are observed among these high-symmetry points. At the $\Gamma$ point, both electrons and holes exhibit relatively weak in-plane anisotropy, characterized by nearly circular hexagonal velocity contours. The Fermi velocity of the VBM ranges from 6.186×10$^3$ m/s at 30° to 6.500×10$^3$ m/s at 0°, while that of the CBM varies from 1.117×10$^3$ m/s at 30° to 1.289×10$^3$ m/s at 0°. At the K point, both electron and hole Fermi velocity distributions display pronounced in-plane anisotropy, forming three-petal patterns with the same orientation, reflecting the underlying lattice symmetry. The Fermi velocity of the VBM exhibits a minimum of 0.406×10$^3$ m/s at 0° and a maximum of 5.867×10$^3$ m/s at 60°, while that of the CBM ranges from 0.084×10$^3$ m/s at 0° to 1.250×10$^3$ m/s at 60°. At the M point, the electron and hole Fermi velocity distributions overlap spatially and exhibit an identical two-lobed (bilobed) annular geometry. The velocities are strongly suppressed, with the VBM varying from 0.002×10$^3$ m/s at 30° to 6.289×10$^3$ m/s at 90°, and the CBM from 0.002×10$^3$ m/s at 30° to 1.153×10$^3$ m/s at 90°. Overall, these results demonstrate that the Fermi velocities are strongly suppressed within the kx–ky plane, even smaller than those reported in twisted bilayer graphene [21-22], consistent with quasi-flat band edges. The pronounced anisotropy at the K point and the overlapping anisotropy at the M point together highlight the momentum-dependent nature of carrier dynamics in twisted diamond. Such symmetry-selective anisotropy provides a real-space manifestation of the extended quasi-flat band manifolds in momentum space and offers a microscopic explanation for the emergence of the 2D–2D domain-direct band gap and its strongly directional transport characteristics.

Notably, the 2D–2D domain-direct band gap identified in structure 165-13-156-r567-0 is not an isolated case. Based on our previously generated high-throughput screening of twisted diamond structures [31], we identify 100 configurations in which both the VBM and CBM reside within the kz = 0 2D BZ and form a direct band gap. Following criteria commonly adopted for quasi-direct gaps (energy differences of  about 0.1–0.12 eV) [32] and flat-band dispersions (energy fluctuations of  about 50 meV) [29, 33], we adopt the more stringent flatness requirement and define domain-direct semiconductors as those in which the energy variations of both the VBM and CBM are smaller than 30 meV across the 2D BZ. Under this criterion, the majority of the identified structures fall into the 2D–2D category, as shown in Figs. S4–S12. When the flatness threshold is further tightened to 20 meV and even 10 meV, the momentum-space extent of the CBM and VBM manifolds is progressively reduced: the corresponding sets Sc and Sv shrink and, in some cases, fragment into symmetry-related patches within the 2D BZ. This smooth evolution indicates that the domain-direct character is continuously tunable rather than a binary classification. Importantly, the representative structure 165-13-156-r567-0 remains exceptional in that, even under the stringent 10 meV criterion, both the CBM and VBM can still be regarded as nearly flat manifolds spanning the entire 2D BZ, as illustrated by the enlarged view in Fig. S13. This places 165-13-156-r567-0 at the extreme limit of the 2D–2D domain-direct regime and establishes it as a prototypical platform for exploring both the fundamental physics and potential applications associated with domain-direct band gaps. Overall, these results demonstrate that domain-direct band gaps in twisted diamond are not rare anomalies, but rather a robust and statistically prevalent feature that can be continuously tuned through structural modulation.

The dual flat-band structure, in which both the CBM and VBM exhibit energy fluctuations within 10 meV across the entire 2D BZ, is expected to have important implications for optical responses. To examine this effect, we calculated the optical absorption spectra of a representative structure, 165-13-156-r567-0, and compared it with bulk diamond and two additional large-size configurations, 163-13-148-r6-0 and 159-26-148-r567-3, as shown in Fig. 5. Owing to the large real-space sizes of all these systems, the optical properties were evaluated using the PBE functional, which is sufficient for capturing qualitative trends in absorption behavior. As reference cases, in the 163-13-148-r6-0 structure (Fig. S14), the CBM remains nearly flat across the 2D BZ with fluctuations below 10 meV, whereas the VBM exhibits noticeable dispersion in a limited momentum region. In contrast, for the 159-26-148-r567-3 structure (Fig. S15), the VBM is nearly flat, while the CBM shows appreciable fluctuations. Strikingly, the structure 165-13-156-r567-0, in which both the CBM and VBM are simultaneously flat over the entire 2D BZ, displays a pronounced optical absorption peak immediately at the band-gap onset, indicating a sharp interband transition. By contrast, both diamond and the two partially flat-band configurations exhibit a gradual increase in absorption above the gap, without a distinct peak feature. These results demonstrate that the coexistence of flat conduction and valence bands plays a crucial role in enhancing optical absorption, as it gives rise to a strongly increased joint density of states at the band edge. This finding highlights the potential of 2D–2D domain-direct semiconductors for optoelectronic applications, where sharp and intense absorption features are highly desirable.

 In this study, we introduce the concept of domain-direct band gaps, a significant extension to the traditional classification of direct band-gap semiconductors. Unlike conventional direct gaps that are defined by point-like band extrema, domain-direct gaps are characterized by extended manifolds in momentum space, where the conduction band minimum (CBM) and valence band maximum (VBM) form quasi-flat surfaces in the Brillouin zone. Our detailed investigation of twisted diamond structures reveals a prototypical example of a 2D–2D domain-direct band gap, with both the CBM and VBM exhibiting near-zero dispersion in the 2D Brillouin zone, resulting in a direct band gap of 3.264 eV. Notably, the flatness of the band edges enhances the joint density of states, leading to a sharp optical absorption peak at the gap edge, unlike other conventional structures that show gradual absorption. Through high-throughput screening of a broad range of twisted diamond configurations, we identify 100 structures exhibiting domain-direct gaps, confirming the prevalence and robustness of this phenomenon across different material variations. The interplay between flat band edges and strong anisotropy in carrier dynamics, combined with the enhanced optical transitions, positions domain-direct semiconductors as promising candidates for novel optoelectronic applications. Our findings not only provide a new classification for semiconductors but also open up exciting possibilities for engineered materials with tunable electronic and optical properties, pushing the boundaries of current semiconductor technology.

\textbf{Acknowledgments}

This work is supported by the Research Foundation of Education Bureau of Hunan Province, China (Grant No. 24A0121) and the National Natural Science Foundation of China (Grant Nos. 52372260, and 12204397, 12374046), the Youth Science and Technology Talent Project of Hunan Province (Grant No. 2022RC1197) and, the Science Fund for Distinguished Young Scholars of Hunan Province of China (No.2024JJ2048).

\textbf{References}

[1] N. W. Ashcroft and N. D. Mermin, Solid State Physics (Holt, Rinehart and Winston, New York, 1976).

[2] C. Kittel, Introduction to Solid State Physics, 8th ed. (Wiley, New York, 2005).

[3] P. Y. Yu and M. Cardona, Fundamentals of Semiconductors: Physics and Materials Properties, 4th ed. (Springer, Berlin, 2010).

[4] D. Leykam, S. Flach, O. Bahat-Treidel, and A. S. Desyatnikov, Flat-band states: Disorder and nonlinearity, Phys. Rev. B 88, 224203 (2013).

[5] S. Mukherjee, A. Spracklen, D. Choudhury, N. Goldman, P. Öhberg, E. Andersson, and R. R. Thomson, Observation of a localized flat-band state in a photonic Lieb lattice, Phys. Rev. Lett. 114, 245504 (2015).

[6] H. Tian, X. Gao, Y. Zhang, S. Che, T. Xu, P. Cheung, K. Watanabe, T. Taniguchi, M. Randeria, F. Zhang, C. Lau, and M. Bockrath, Evidence for Dirac flat-band superconductivity enabled by quantum geometry, Nature 614, 440 (2023).

[7] G. Sethi, Y. Zhou, L. Zhu, L. Yang, and F. Liu, Flat-band-enabled triplet excitonic insulator in a diatomic kagome lattice, Phys. Rev. Lett. 126, 196403 (2021).

[8] I. Hase, T. Yanagisawa, Y. Aiura, and K. Kawashima, Possibility of flat-band ferromagnetism in hole-doped pyrochlore oxides Sn$_2$Nb$_2$O$_7$ and Sn$_2$Ta$_2$O$_7$, Phys. Rev. Lett. 120, 196401 (2018).

[9] C. He, Y. Liao, T. Ouyang, H. Zhang, H. Xiang, and J. Zhong, Flat-band-based high-temperature ferromagnetic semiconducting state in the graphitic C$_4$N$_3$ monolayer, Fundam. Res. 5, 138 (2025).

[10] C. He, S. Li, Y. Zhang, Z. Fu, J. Li, and J. Zhong, Isolated zero-energy flat bands and intrinsic magnetism in carbon monolayers, Phys. Rev. B 111, L081404 (2025).

[11] B. Liu, S. Liu, Y. Zhang, C. He, J. Yang, and H. Xiang, Designing flat-band materials through compact localized state clusters, Phys. Rev. B 112, L241111 (2025).

[12] R. Yu, H. Weng, Z. Fang, X. Dai, and X. Hu, Topological nodal-line semimetal and Dirac semimetal state in antiperovskite Cu$_3$PdN, Phys. Rev. Lett. 115, 036807 (2015).

[13] H. Weng, Y. Liang, Q. Xu, R. Yu, Z. Fang, X. Dai, and Y. Kawazoe, Topological nodal-line semimetal in three-dimensional graphene networks, Phys. Rev. B 92, 045108 (2015).

[14] A. A. Burkov, M. D. Hook, and L. Balents, Topological nodal semimetals, Phys. Rev. B 84, 235126 (2011).

[15] N. P. Armitage, E. J. Mele, and A. Vishwanath, Weyl and Dirac semimetals in three-dimensional solids, Rev. Mod. Phys. 90, 015001 (2018).

[16] T. Bzdušek, Q. Wu, A. Rüegg, M. Sigrist, and A. A. Soluyanov, Nodal-chain metals, Nature 538, 75 (2016).

[17] M. Z. Hasan, S.-Y. Xu, I. Belopolski, and S.-M. Huang, Discovery of Weyl fermion semimetals and topological Fermi arc states, Annu. Rev. Condens. Matter Phys. 8, 289 (2017).

[18] E. Tang, J.-W. Mei, and X.-G. Wen, High-temperature fractional quantum Hall states, Phys. Rev. Lett. 106, 236802 (2011).

[19] L. Balents, C. R. Dean, D. K. Efetov, and A. F. Young, Superconductivity and strong correlations in moiré flat bands, Nat. Phys. 16, 725 (2020).

[20] R. S. K. Mong, A. M. Essin, and J. E. Moore, Antiferromagnetic topological insulators, Phys. Rev. B 83, 125109 (2011).

[21] Y. Cao, V. Fatemi, S. Fang, K. Watanabe, T. Taniguchi, E. Kaxiras, and P. Jarillo-Herrero, Unconventional superconductivity in magic-angle graphene superlattices, Nature 556, 43 (2018).

[22] Y. Cao, V. Fatemi, A. Demir, S. Fang, S. L. Tomarken, J. Y. Luo, J. D. Sanchez-Yamagishi, K. Watanabe, T. Taniguchi, E. Kaxiras, R. C. Ashoori, and P. Jarillo-Herrero, Correlated insulator behaviour at half-filling in magic-angle graphene superlattices, Nature 556, 80 (2018).

[23] E. Codecido, Q. Wang, R. Koester, S. Che, H. Tian, R. Lv, S. Tran, K. Watanabe, T. Taniguchi, F. Zhang, M. Bockrath, and C. N. Lau, Correlated insulating and superconducting states in twisted bilayer graphene below the magic angle, Sci. Adv. 5, eaaw9770 (2019).

[24] X. Wang, C. Xu, S. Aronson, D. Bennett, N. Paul, P. J. D. Crowley, C. Collignon, K. Watanabe, T. Taniguchi, R. Ashoori, E. Kaxiras, Y. Zhang, P. Jarillo-Herrero, and K. Yasuda, Moiré band structure engineering using a twisted boron nitride substrate, Nat. Commun. 16, 178 (2025).

[25] C. S. Tsang, X. Zheng, T. Yang, Z. Yan, W. Han, L. W. Wong, H. Liu, S. Gao, K. H. Leung, C. S. Lee, S. P. Lau, M. Yang, J. Zhao, and T. H. Ly, Polar and quasicrystal vortex observed in twisted-bilayer molybdenum disulfide, Science 386, 198 (2024).

[26] J. Cai, E. Anderson, C. Wang, X. Zhang, X. Liu, W. Holtzmann, Y. Zhang, F. Fan, T. Taniguchi, K. Watanabe, Y. Ran, T. Cao, L. Fu, D. Xiao, W. Yao, and X. Xu, Signatures of fractional quantum anomalous Hall states in twisted $\mathrm{MoTe_2}$, Nature 622, 63 (2023).

[27] L. Wang, E. M. Shih, A. Ghiotto, L. Xian, D. A. Rhodes, C. Tan, M. Claassen, D. M. Kennes, Y. Bai, B. Kim, K. Watanabe, T. Taniguchi, X. Zhu, J. Hone, A. Rubio, A. N. Pasupathy, and C. R. Dean, Correlated electronic phases in twisted bilayer transition metal dichalcogenides, Nat. Mater. 19, 861 (2020).

[28] X. Shi, C. He, C. J. Pickard, C. Tang, and J. Zhong, Stochastic generation of complex crystal structures combining group and graph theory with application to carbon, Phys. Rev. B 97, 014104 (2018).

[29] S. Li, X. Shi, J. Li, C. He, T. Ouyang, C. Tang, and J. Zhong, Spin-orbital coupling induced isolated flat bands in bismuthene with k-dependent spin texture, Phys. Rev. B 110, 115115 (2024).

[30] Y. Wei, S. Li, X. Shi, and C. He, Low-energy tetrahedral networks for carbon and silicon from (2+1)-regular bipartite-like graphs, IUCrJ 12, 523 (2025).

[31] Y Wei, S Li, Y Song, C He, Two-dimensional flat-bands in Moire-diamonds, arXiv:2510.10908

[32] H. Liu, S. Meng, and F. Liu, Screening two-dimensional materials with topological flat bands, Phys. Rev. Mater. 5, 084203 (2021).

[33] I.-H. Lee, J.-Y. Lee, Y.-J. Oh, S.-Y. Kim, and K. J. Chang, Ab initio structure search and stability of carbon allotropes, Phys. Rev. B 90, 115209 (2014).

\end{document}